\documentclass[%
superscriptaddress,
nofootinbib,
 amsmath,amssymb,
 twocolumn,
 prl,
pra,
floatfix,
]{revtex4-1}

\usepackage{xcolor}
\usepackage{graphicx}
\usepackage{dcolumn}
\usepackage{bm}
\usepackage{hyperref}
\usepackage{multirow}

\begin{document}

\title{Power Corrections to Energy Flow Correlations from Large Spin Perturbation}

\author{Hao Chen}
 \email{chenhao201224@zju.edu.cn}
 \affiliation{Zhejiang Institute of Modern Physics, School of Physics, Zhejiang University, Hangzhou, 310027, China}
 
\author{Xinan Zhou}
 \email{xinan.zhou@ucas.ac.cn}
\affiliation{Kavli Institute for Theoretical Sciences,
University of Chinese Academy of Sciences, Beijing 100190, China.
}%

\author{Hua Xing Zhu}
 \email{zhuhx@zju.edu.cn}
\affiliation{Zhejiang Institute of Modern Physics, School of Physics, Zhejiang University, Hangzhou, 310027, China}

\begin{abstract}
Dynamics of high energy scattering in Quantum Chromodynamics~(QCD) are primarily probed through detector energy flow correlations. One important example is the Energy-Energy Correlator~(EEC), whose back-to-back limit probes correlations of QCD on the lightcone and can be described by a transverse-momentum dependent factorization formula in the leading power approximation. In this work, we develop a systematic method to go beyond this approximation. We identify the origin of logarithmically enhanced contributions in the back-to-back limit as the exchange of operators with low twists and large spins in the local operator product expansion.  Using techniques from the conformal bootstrap, the large logarithms beyond leading power can be resummed to all orders in the perturbative coupling. As an illustration of this method, we perform an all-order resummation of the leading and next-to-leading logarithms beyond the leading power in ${\cal N} = 4$ Super Yang-Mills theory. 
\end{abstract}

\maketitle

\section{Introduction}

Distributions of energy flows in high energy scattering in Quantum Chromodynamics~(QCD) encode unique information of the Lorentizian dynamics of quantum gauge theory, which is otherwise hard to extract. A famous example is the formation of jets, which are collimated sprays of hadrons that manifest the underlying structure of of quark and gluon scattering~\cite{Hanson:1975fe,Sterman:1977wj}. From the early days of QCD, global energy flow distributions have been characterized using infrared and collinear safe shape functions~\cite{Bjorken:1969wi,Ellis:1976uc,Georgi:1977sf,Farhi:1977sg,Parisi:1978eg,Donoghue:1979vi,Rakow:1981qn,Berger:2003iw,Stewart:2010tn}. Alternatively, it can also be studied using statistical correlation of final-state energy flows, of which the simplest one is the two-point Energy-Energy Correlator~(EEC)~\cite{Basham:1978bw,Basham:1977iq,Basham:1979gh,Basham:1978zq}. In fact, the two different approaches for analyzing energy flow distributions are closely related by an integral transformation~\cite{Belitsky:2001ij,Sterman:2004pd,Komiske:2017aww,Chen:2020vvp}. 

In the $e^+e^-$ scattering, EEC can be written as the Fourier transformation of the correlation function of energy flow operators 
\begin{equation}
\label{eq:def}
    \text{EEC}(y) =\frac{8\pi^2}{q^2 \sigma_{0}} \! \int \! d^4 x\, e^{i q \cdot x_{13}} \langle    J^\mu(x_1)  {\cal E}(n_2) {\cal E}(n_4) J_\mu^\dagger(x_3) \rangle \,,
\end{equation}
where $n_2 = (1, \vec{n}_2)$ and $n_4 = (1, \vec{n}_4)$ specify the directions of the detected energy flow in the collider, $y = 1-\frac{(n_2\cdot n_4) q^2}{2(n_2\cdot q)(n_4\cdot q)} =(1 + \cos\theta)/2$, $q^2>0$ is a timelike momentum, $J^\mu$ is the electromagnetic current, and the energy flow operator is defined as a detector time integral of the energy-momentum tensor~\cite{Sveshnikov:1995vi,Tkachov:1995kk,Korchemsky:1999kt,Bauer:2008dt,Hofman:2008ar,Belitsky:2013xxa,Belitsky:2013bja,Kravchuk:2018htv}
\begin{equation}
\mathcal{E}(n_i)= \int \displaylimits_{-\infty}^\infty \frac{d\, n_i \! \cdot \! x_i}{16} \lim_{\bar{n}_i \! \cdot \! x_i \to \infty} (\bar{n}_i\cdot x_i)^2 T_{\mu\nu}(x_i)\bar{n}_i^\mu \bar{n}_i^\nu\,.
\end{equation}
Experimental studies of EEC have a long history~\cite{SLD:1994idb,L3:1992btq,OPAL:1991uui,TOPAZ:1989yod,TASSO:1987mcs,JADE:1984taa,Fernandez:1984db,Wood:1987uf,CELLO:1982rca,PLUTO:1985yzc}. They have been used to provide precision extraction of strong coupling constant~\cite{deFlorian:2004mp,Kardos:2018kqj}.

Intuitively, when two narrow jets are produced in $e^+e^-$, they tend to be back-to-back due to momentum conservation, reflecting the underlying $q\bar q$ production. This leads to a peak for EEC as $y \to 0$, known as the Sudakov peak. In perturbation theory, it is manifested as large logarithmic corrections,
\begin{equation}
\label{eq:schematic}
\text{EEC}(y) \sim \sum_{n = 1}^\infty \sum_{m=0}^{2n-1}\alpha_s^n 
 \left( c_{n,m}\frac{\log^{m} y}{y} + d_{n,m}\log^m y\right) \,,
\end{equation}
where we have only shown the Leading Power~(LP, $\sim y^{-1}$) and the Next-to-Leading Power~(NLP, $\sim y^0$) corrections for simplicity.
The leading power series of EEC resembles the perturbative structure of vector boson production at small $p_T$ and exhibits lightcone divergences. It can be resummed to all orders in the perturbative coupling by solving a 2d renormalization group equation in virtuality and rapidity~\cite{Collins:1981uk,Moult:2018jzp}.  Using the recently available 4-loop rapidity anomalous dimension~\cite{Moult:2022xzt,Duhr:2022yyp}, its perturbative resummation has been performed to N$^4$LL accuracy~\cite{Duhr:2022yyp}.
However, power corrections to EEC, and in general to event shape functions and transverse momentum dependent observables, are much less understood, both perturbatively and non-perturbatively. Recently there have been significant developments towards a more satisfying picture of power corrections for various observables, see {\it e.g.}~\cite{Pirjol:2002km,Beneke:2002ph,Bonocore:2014wua,Bonocore:2015esa,Moult:2016fqy,Boughezal:2016zws,Feige:2017zci,DelDuca:2017twk,Moult:2017rpl,Moult:2017jsg,Goerke:2017lei,Beneke:2017ztn,Moult:2018jjd,Ebert:2018lzn,Beneke:2018rbh,Boughezal:2018mvf,Ebert:2018gsn,Moult:2019mog,vanBeekveld:2019prq,Bahjat-Abbas:2019fqa,Bacchetta:2019qkv,Cieri:2019tfv,Buonocore:2019puv,Moult:2019uhz,Beneke:2019oqx,Moult:2019vou,Liu:2019oav,Ebert:2020dfc,Beneke:2020ibj,Luisoni:2020efy,Inglis-Whalen:2020rpi,Inglis-Whalen:2021bea,Liu:2020tzd,Liu:2020wbn,vanBeekveld:2021hhv,vanBeekveld:2021mxn,Salam:2021tbm,Caola:2021kzt,Ebert:2021jhy,Vladimirov:2021hdn,Buonocore:2021tke,Camarda:2021jsw,Luke:2022ops,Liu:2022ajh,Gamberg:2022lju}. Yet the status is still far from that of leading power terms.  

In this work we initiate a study of power corrections to EEC by exploiting conformal symmetry and techniques from the analytic conformal bootstrap~\cite{Rychkov:2016iqz,Simmons-Duffin:2016gjk,Poland:2018epd,Bissi:2022mrs}.\footnote{Applications of conformal symmetry in QCD has a long history~\cite{Braun:2003rp}. For recent applications, see {\it e.g.} \cite{Vladimirov:2016dll,Braun:2017cih,Braun:2019wyx,Chen:2022jhb,Chang:2022ryc,Moult:2022xzt,Duhr:2022yyp}.}
Building upon an important observation by Korchemsky~\cite{Korchemsky:2019nzm}, we connect the logarithmically enhanced terms in power corrections~($m>0$ in \eqref{eq:schematic}) with the expansion of correlators around the double lightcone limit, which is controlled by the twist expansion and the large spin expansion. Using twist conformal blocks \cite{Alday:2016njk}, tails from large spins can be systematically resummed. Further simplifications come from crossing symmetry, which relates the twist corrections with large spin corrections. Using this method, we explicitly carry out a calculation in ${\cal N} = 4$ super Yang-Mills (SYM) theory, and achieve the first Leading and Next-to-Leading Logarithmic resummation at the subleading power.

\section{Back-to-back v.s. double lightcone}

\begin{figure}
    \centering
    \includegraphics[width = 0.4\textwidth]{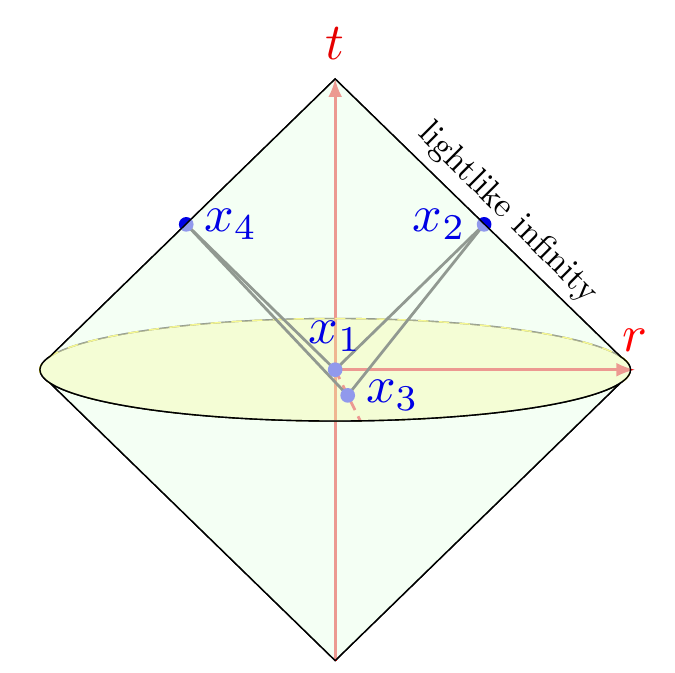}
    \caption{Penrose diagram for the double lightcone limit of the local correlator.}
    \label{fig:penrose}
\end{figure}

As is clear from \eqref{eq:def}, EEC is related to the local Wightman correlator $\langle \Omega |J(x_1) T(x_2) T(x_4) J(x_3)| \Omega \rangle$ by detector time integrals and a Fourier transform~\cite{Belitsky:2013xxa,Belitsky:2013bja}. It is useful to understand which region of the local correlator corresponds to the $y \to 0$ limit of EEC. Let us backtrack the dominant contribution by first undoing the Fourier transformation. We can apply a Lorentz transformation to make the two detectors exactly back-to-back: $n_4^\mu\to \bar{n}_2^\mu = (1,-\vec{n}_2)$. At the same time, the momentum $q^\mu$ gains a small transverse component $\vec{q}^\perp$, which is schematically related to $y$ as $|\vec{q}^\perp|^2/q^2\sim y$. In position space, this corresponds to the region where $|(n_2\cdot x_{13})(\bar{n}_2\cdot x_{13})/x_{13}^2|\sim y$. Since $\mathcal{E}(n_2)\mathcal{E}(\bar n_2)$ is invariant under the boost along $\vec{n}_2$, we have an extra degree of freedom to choose a frame such that $n_2\cdot x_{13},\bar{n}_2\cdot x_{13}\sim \sqrt{y}|x_{13}|$. In such a frame, the lightcone singularities of $1$ and $3$ are very close on the path of detector time integrals of $2$ and $4$. Therefore, we expect the dominant contribution to come from the region where $2$ and $4$ are near the pinch of lightcone singularities: $x_{12}^2, x_{23}^2, x_{34}^2, x_{14}^2\ll x_{13}^2, x_{24}^2$, which is the null square configuration~\cite{Alday:2010zy}. Instead, we can also choose the frame that $n_2\cdot x_{13}\sim y |x_{13}|, \bar{n}_2\cdot x_{13}\sim |x_{13}|$. Then the leading $y\to 0$ dependence comes from 2 being near the lightcone limit of both $1$ and $3$, which is called the {\it double lightcone limit}. In this work we delineate this correspondence and systematically go beyond the leading power results in \cite{Korchemsky:2019nzm}. 

It is well known that the lightcone limit is controlled 
 by twist expansion~\cite{Ferrara:1971zy,Ferrara:1972cq,Christ:1972ms,Muta:1998vi}.
The most singular behavior in the $1,2$-channel as $x_{12}^2\to 0$ is produced by operators with the lowest  twist $\tau=\Delta-\ell$. However, any single operator in the $1,2$-channel cannot generate the lightcone singularity in the crossed channel with $x_{13}^2\to 0$. The $x_{13}^2\to 0$ singularity can only be produced by an infinite sum of operators with different spins at a given twist. Therefore, the double lightcone limit, or relatedly the back-to-back limit of EEC, is determined by the large spin asymptotics of low-twist operators.

To be specific, we now consider EEC in ${\cal N} = 4$ SYM. Comments for QCD will be given at the end. We consider the correlator of scalar operators belonging to the stress tensor supermultiplet
\begin{equation}\label{4ptfun}
    \langle \mathcal{O}(x_1) \mathcal{O}(x_2) \mathcal{O}(x_3) \mathcal{O}(x_4) \rangle_{\rm dyn} = \frac{1}{(2\pi)^4}\frac{x_{13}^4x_{24}^4}{(x_{12}^6 x_{34}^2)^6}\mathcal{F}(u,v)\,.
\end{equation}
Here we have only kept the dynamical part and subtracted the contribution of protected operators which are not perturbatively corrected, see Supplemental Material for details.
Conformal symmetry ensures that $\mathcal{F}$ is a function of the conformal cross ratios 
\begin{equation}
u=\frac{x_{12}^2 x_{34}^2}{x_{13}^2 x_{24}^2}=z\bar{z}\;,\quad v=\frac{x_{23}^2 x_{14}^2}{x_{13}^2 x_{24}^2}=(1-z)(1-\bar{z})\;.
\end{equation}
Expanded in the small coupling $a=\frac{g^2 N_c}{4\pi^2}$, 
$\mathcal{F}(u,v)$ reads
\begin{equation}
\mathcal{F}(u,v)=\sum_{n=0}^{\infty}a^n \mathcal{F}^{(n)}(u,v)
=\mathcal{F}^{(0)}(u,v)+\frac{u^3}{v} \Phi(u,v)\,,
\end{equation}
The new function $\Phi(u,v)=\sum_{n\geq 1}a^n \Phi^{(n)}(u,v)$ packages all the coupling dependent information and is crossing symmetric, {\it i.e.}, $\Phi(u,v)=\Phi(v,u)$. It has been calculated to three loops in~\cite{Eden:2011we,Drummond:2013nda}. The correlator admits the following superconformal block decomposition \cite{Dolan:2001tt,Beem:2016wfs}:
\begin{equation} \label{eq: super_blocks}
    \mathcal{F}(u,v)=\sum_{\Delta}\sum_{\text{even } \ell} a_{\tau,\ell} G_{\Delta+4,\ell}(u,v)\,.
\end{equation}
Here the sums are over all superconformal primary operators with dimension $\Delta$, spin $\ell$ and OPE coefficient $a_{\tau,\ell}$. The use of $\tau$ in the label foreshadows the twist expansion later. The 4d {\it bosonic} conformal blocks are given by \cite{Dolan:2000ut}
\begin{equation}
    G_{\Delta,\ell}(u,v)=\frac{z\bar{z}}{\bar{z}-z}\left[k_{\Delta-\ell-2}(z)k_{\Delta+\ell}(\bar{z})-(z\leftrightarrow \bar{z})\right]\,,
\end{equation}
where $k_\beta(x)=x^{\beta/2}{}_2F_1(\frac{\beta}{2},\frac{\beta}{2},\beta;x)$. They are eigenfunctions of the quadratic Casimir operator 
\begin{equation}
\mathcal{D}_2= z^2((1-z)\partial_z^2-\partial_z)+
\frac{(d-2)z\bar{z}}{z-\bar{z}}(1-z)\partial_z+(z\leftrightarrow \bar{z})
\end{equation}
with eigenvalues $\frac{1}{2}\left(\Delta(\Delta-4)+\ell(\ell+2)\right)$ \cite{Dolan:2003hv}. Superconformal symmetry causes a shift $\Delta\to\Delta+4$ in the expansion (\ref{eq: super_blocks}), and we  denote $\mathbf{G}_{\tau,\ell}(z,\bar{z})=G_{\Delta+4,\ell}(u,v)$ for convenience. Perturbative corrections, via conformal block decomposition, are encoded in the expansions of twists and OPE coefficients
\begin{equation}
\tau=\tau_0+\sum_{n=1}^{\infty}a^n \gamma^{(n)}_{\tau_0, \ell}\,,\quad
    a_{\tau,\ell}= \sum_{n=0}^\infty a^n a^{(n)}_{\tau_0,\ell}\,,
\end{equation}
where $\tau_0$ is the classical twist and $\sum_{n\geq 1} a^n \gamma^{(n)}_{\tau_0,\ell}=\gamma_{\tau,\ell}$ is the anomalous dimension. The anomalous dimensions enter via the expansion for conformal blocks: $\mathbf{G}_{\tau,\ell}=\sum_{n= 0}^{\infty}\frac{1}{n!}\gamma_{\tau,\ell}^n \partial_{\tau_0}^n \mathbf{G}_{\tau_0,\ell}$. Note that $\partial_{\tau_0}^n \mathbf{G}_{\tau_0,\ell}$ contains at most $\log^n u$ in the small $u$ limit.

The double lightcone limit corresponds to $u, v\to 0$, or equivalently, $z\to 0,\, \bar{z}\to 1$. Each conformal block in this limit is controlled by the twist $G_{\Delta,\ell}(z,\bar{z})\sim z^{\tau/2} \log(1-\bar{z})$. 
The presence of logarithms partially demonstrates the effect of summing over infinitely many spinning operators because each conformal block contains infinitely many descendants in a given conformal family. However, we will encounter more divergent pieces than a single logarithm, which require the large spin contribution from infinitely many {\it primary} operators. Examples include power divergences $(1-\bar{z})^{m<0}$ and powers of logarithms $[\log(1-\bar{z})]^{k\geq 2}$~\footnote{At one loop the perturbative logarithms at NLP have at most $k=1$, therefore are not fixed by large spins.}. The latter arises from $\log^k u$ of $\partial_{\tau_0}^k \mathbf{G}_{\tau_0,\ell}$ in small $u$ under crossing. In the following, we refer to these as {\it enhanced divergences}. To lighten the notations, we will denote the logarithms as $\log(x)=L_x$ from now on.

We also point out that crossing symmetry will play an important role in our computation of the power corrections. As we will see, the $u$, $v$ power corrections to $\Phi(u,v)$ contribute equally to the $y$ power corrections in EEC. In particular, EEC at NLP corresponds to order $u^0 v^1$ and $u^1 v^0$ in $\Phi(u,v)$. The former only requires twist-2 data to subleading order in the large spin limit~($\mathcal{O}(\ell^{-2})$), while the latter contains twist-4 contributions.  Crossing symmetry equates these two contributions and therefore avoids the necessity of inputting the twist-4 information. 

\begin{table}[htp]
\caption{CFT data needed at different orders}
\begin{center}
\def\arraystretch{1.5}
\begin{tabular}{|c|c|c|c|c|}
\hline
\multirow{2}{*}{} &\multicolumn{2}{c|}{ Power Corrections} & \multicolumn{2}{c|}{ Perturbative Corrections}\\ \cline{2-5}
 & twist & large spin & LL & NLL \\ \hline
 LP & 2 & $\mathcal{O}(\ell^0)$ & $a_{2,\ell}^{(0)}\,,  \gamma_{2,\ell}^{(1)}$ & $a_{2,\ell}^{(1)}\,,  \gamma_{2,\ell}^{(2)}$\\ \hline
 \multirow{2}{*}{NLP} & 2 & $\mathcal{O}(\ell^{-2})$ & $a_{2,\ell}^{(0)}\,,  \gamma_{2,\ell}^{(1)}$ & $a_{2,\ell}^{(1)}\,,  \gamma_{2,\ell}^{(2)}$\\\cline{2-5}
 & 4 & $\mathcal{O}(\ell^0)$ & $a_{4,\ell}^{(0)}\,,  \gamma_{4,\ell}^{(1)}$ & $a_{4,\ell}^{(1)}\,,  \gamma_{4,\ell}^{(2)}$\\ \hline
\end{tabular}
\end{center}
\label{Tab:data}
\end{table}%

Before going into the technical details, we provide a brief overview for the next two sections. We will first use techniques from large spin perturbation theory to extract the enhanced divergences in $v$ at twist-2 up to NLP. That is, we will obtain $\Phi(u,v)= p_0(L_u, L_v) + p_1(L_u,L_v) v +\cdots$, in which $p_0$, $p_1$ are polynomials in $L_u$, $L_v$ at each perturbative order. Crossing symmetry then fixes the NLP contribution in $u$ to be $\Phi(u,v)= p_0(L_u, L_v) + p_1(L_u,L_v) v +p_1(L_v,L_u) u\cdots$. The relevant data is listed in Table \ref{Tab:data}, but only the first two rows are needed thanks to crossing symmetry. Finally, we map the small $u$, $v$ expansion of $\Phi(u,v)$ to the back-to-back limit of $\mathrm{EEC}(y)$. The rules  at LP and NLP are given in Table \ref{tab: log_map} and are valid to NLL accuracy. We will explain these points in more detail in the rest of the paper.

\begin{table}[htp]
\caption{Logarithms Map at LP and NLP}
\begin{center}
\def\arraystretch{1.6}
\begin{tabular}{|c|c|c|}
\hline 
 & $\Phi(u,v)$ & $4 y (1-y)^2 \times\mathrm{EEC}(y)$\\
 \hline
 \hline
 LP & $L_u^m L_v^m$ & $2(m+n) L^{m+n-1}_{y/(1-y)}$\\
 \hline
 \multirow{2}{*}{ NLP} & $u L_u^m L_v^n$ & $\frac{2m(1-m)}{m+n-1} \frac{y}{1-y} L^{m+n-1}_{y/(1-y)}$ \\ \cline{2-3}
 & $v L_u^m L_v^n$ &  $\frac{2n(1-n)}{m+n-1} \frac{y}{1-y} L^{m+n-1}_{y/(1-y)}$ \\ \hline
\end{tabular}
\end{center}
\label{tab: log_map}
\end{table}%

\section{Large spin analysis}
The enhanced divergences can be systematically handled by using the Large Spin Perturbation Theory~\cite{Alday:2016njk}, which culminates an array of earlier works in the large spin sector \cite{Alday:2007mf,Fitzpatrick:2012yx,Komargodski:2012ek,Alday:2013cwa,Alday:2015eya,Alday:2015ota,Alday:2015ewa}. The starting point is the free theory limit where the twists are degenerate. The correlators can be written as a sum over the twists
\begin{equation}
\mathcal{F}^{(0)}(z,\bar{z})=\sum_{\tau_0=2,4,\ldots} H_{\tau_0}(z,\bar{z})\;,
\end{equation}
where each $H_{\tau_0}(z,\bar{z})$ sums over spins
\begin{equation}
H_{\tau_0}(z,\bar{z})=\sum_{\ell=0}^\infty \langle a_{\tau_0,\ell}^{(0)}\rangle \mathbf{G}_{\tau_0,\ell}(z,\bar{z})\;,
\end{equation}
and is known as a {\it twist conformal block} (TCB). When the interaction is turned on, the twist degeneracies are lifted and the OPE coefficients are corrected. In general, we find that the expansion consists of sums of the form
\begin{equation}
\sum_{\ell=0}^\infty \langle a_{\tau_0,\ell}^{(0)}\rangle \kappa_{\tau_0}(\ell) \mathbf{G}_{\tau_0,\ell}(z,\bar{z})\;,
\end{equation}
where the quantities $\kappa_{\tau_0}(\ell)$ admit expansions around large conformal spins $J^2_{\tau,\ell}=(\ell+\frac{\tau}{2})(\ell+\frac{\tau}{2}-1)$ with the following schematic form
\begin{equation}
\kappa_{\tau_0}(\ell)=\sum_{m=0}^\infty\sum_{i=0}^N \frac{K_{m,i}}{J_{\tilde \tau_0,\ell}^{2m}}\log^i J_{\tilde\tau_0,\ell}^2\;.
\end{equation}
Here we introduce shifted twists $\tilde\tau_0=\tau_0+4$ to take into account the dimension shift in the decomposition (\ref{eq: super_blocks}).
An interesting feature, as we will see in explicit calculations, is that only {\it even} negative powers appear in the expansion. This is closely related to the reciprocity relation \cite{Dokshitzer:2005bf,Basso:2006nk}, which has been explicitly verified in QCD to three loops~\cite{Chen:2020uvt}. Consequently, the correlator should now be expanded in terms of a more general class of TCBs 
\begin{equation}
H_{\tau_0}^{(m, i)}(z,\bar{z}) = \sum_{\ell} a^{(0)}_{\tau_0,\ell}\frac{\log^i J^2_{\tilde\tau_0,\ell}}{J_{\tilde \tau_0,\ell}^{2m}} \mathbf{G}_{\tau_0,\ell}(z,\bar{z})\;.
\end{equation}
Note that conformal blocks satisfy the Casimir equation
\begin{equation}
\mathcal{C}_{\tilde \tau_0} \mathbf{G}_{\tau_0,\ell}(z,\bar{z})= J_{\tilde \tau_0,\ell}^2 \mathbf{G}_{\tau_0,\ell}(z,\bar{z})\;,
\end{equation}
where $\mathcal{C}_\tau=\mathcal{D}_2+\frac{1}{4}\tau(2d-\tau-2)$ is the shifted conformal Casimir.
It follows that TCBs obey the following recursion relations
\begin{equation}\label{ConHfull}
H_{\tau_0}^{(m, i)}(z,\bar{z}) = \mathcal{C}_{\tilde \tau_0} H_{\tau_0}^{(m+1, i)}(z,\bar{z})\;. 
\end{equation}
The full TCBs are in general difficult to compute. However, we will only need them in the small $v$ limit where computations become manageable. The TCBs in 4D take a factorized form
\begin{equation} \label{eq: 4d_TCB}
H_{\tau_0}^{(m, i)}(z,\bar{z}) = \frac{z\, k_{\tilde \tau_0-2}(z)}{\bar{z}-z} \bar{H}_{\tau_0}^{(m, i)}(\bar{z})\;,
\end{equation}
where we have dropped the regular part when $\bar{z}\to 1$. Moreover, the recursion relation (\ref{ConHfull}) becomes
\begin{equation}
\bar{H}_{\tau_0}^{(m, i)}(\bar{z}) = \bar{D} \bar{H}_{\tau_0}^{(m+1, i)}(\bar{z})\;,
\end{equation}
where 
\begin{equation}
\bar{D}=\bar{z}^2(1-\bar{z})\frac{d^2}{d\bar{z}^2}
-\bar{z}(2-\bar{z})\frac{d}{d\bar{z}}
+2-\bar{z}\;.
\end{equation}
Using this recursion relation, one can compute the TCBs explicitly in the small $v$ limit ~\cite{Henriksson:2017eej}. For example, we find  
\begin{eqnarray}
\bar{H}_2^{(0,i)}(\bar{z})&=&(-1)^i\left[\frac{L_\epsilon^i}{2\epsilon}+\frac{L_\epsilon^{i+1}}{6(i+1)}+\frac{\gamma_E-3}{3} L_\epsilon^i\right]+\cdots\,,\nonumber\\
\bar{H}_2^{(1,i)}(\bar{z})&=& (-1)^i\left[\frac{L_\epsilon^{i+2}}{2(i+1)(i+2)}+\frac{\gamma_E L_\epsilon^{i+1}}{i+1}\right]+\cdots\,. 
\label{eq: Hb_eg}
\end{eqnarray}
where we have defined $\epsilon=1-\bar{z}$.

Let us now focus on the small $u$ limit, {\it i.e.}, $z\to 0$. Together with $\bar{z}\to1$, the correlator takes the form
\begin{equation}
\begin{split}
\nonumber\mathcal{F}^{(n)}={}&z^3\frac{\log^nz}{\epsilon}\big(\log^n\epsilon(A_{n,1}+A_{n,2}\epsilon+\ldots)\\
{}&+\log^{n-1}\epsilon(B_{n,1}+B_{n,2}\epsilon+\ldots)+\ldots\big)+\mathcal{O}(z^4)\\
={}&z^3\sum_{m=0}^\infty \sum_{i=0}^n C_{m,i} \bar{H}_2^{(m,i)}(\bar{z})+\mathcal{O}(z^4)\;.
\end{split}
\end{equation}
An important point is that to compute the N$^q$LP in the small $\epsilon$ expansion, {\it i.e.}, $A_{n,j=1,2,\ldots,q}$, $B_{n,j=1,2,\ldots,q}$ {\it etc}, we only need $\bar{H}_2^{(m,i)}$ with $m=0,1,\ldots, q$. To see this, we act on the two expansions with $\bar{D}$ and compare the power divergences. Note that for any polynomial $p(\epsilon)$
\begin{equation}
\nonumber \bar{D}(p(\epsilon)\log^i \epsilon)=\frac{i(i-1)(1-\epsilon) p(\epsilon)\log^{i-2}\epsilon}{\epsilon}+\mathcal{O}(\epsilon^0)\;.
\end{equation}
Taking $p(\epsilon)=\epsilon^q$  at the $q$-th order, we find the RHS only becomes a power divergence after acting $q$ times with $\bar{D}$. On the other hand, it is known that only the TCBs with $m\leq 0$ are power divergent. The repeated $\bar{D}$ action makes the TCBs with $m\leq q$ power divergent and therefore responsible for the  $q$-th order correction.

We now  explicitly compute the power corrections using the TCB decomposition. From Table \ref{Tab:data}, to NLL and 2nd order in NLP the needed data is~\cite{Dolan:2004iy, Henriksson:2017eej, Kologlu:2019mfz}
\begin{eqnarray}
    a_{2,\ell}^{(0)}&=& \frac{\Gamma(\ell+3)^2}{\Gamma(2\ell+5)}\,,\\
    \gamma_{2,\ell}^{(1)}&=&\log J^2_{6,\ell}+2\gamma_E+\frac{1}{3 J^2_{6,\ell}}+\mathcal{O}(J^{-4}_{6,\ell})\,,
    \label{eq: large_J_1}\\
    \frac{a_{2,\ell}^{(1)}}{a_{2,\ell}^{(0)}}&=& \! \bigg[\frac{1+4J_{6,\ell}}{16 J^2_{6,\ell}} \!-\! \log 2\bigg]\! \gamma^{(1)}_{2,\ell}\!-\zeta_2\! +\!\frac{1}{J_{6,\ell}}\!+\!\mathcal{O}(J^{-3}_{6,\ell}),
    \label{eq: large_J_2}\\
    \gamma_{2,\ell}^{(2)}&=&\left(\frac{1}{J_{6,\ell}}-\frac{\zeta_2}{2}\right)\gamma_{2,\ell}^{(1)}-\frac{3\zeta_3}{2}+\frac{1}{J^2_{6,\ell}}+\mathcal{O}(J^{-3}_{6,\ell})\,.
    \label{eq: large_J_3}
\end{eqnarray}
From the conformal block decomposition (\ref{eq: super_blocks}), the small $u$ expansion of $\mathcal{F}^{(n)}$ up to NLL accuracy reads
\begin{equation} 
\begin{split} \label{eq: F_twist_NLL_expand}
    \mathcal{F}^{(n)}= z^3\sum_{\text{even }\ell} a^{(0)}_{2,\ell} \Bigg\{ \frac{\left(\gamma^{(1)}_{2,\ell}\right)^n}{2^n n!} L_z^n +\frac{\left(\gamma_{2,\ell}^{(1)}\right)^{n-1} L_z^{n-1} }{2^{n-1}(n-1)!}\\
    \times\left[\frac{a^{(1)}_{2,\ell}}{a^{(0)}_{2,\ell}}+(n-1)\frac{\gamma_{2,\ell}^{(2)}}{\gamma_{2,\ell}^{(1)}}+\frac{\gamma_{2,\ell}^{(1)}\partial_\ell}{2}\right]\Bigg\} k_{2\ell+6}(\bar{z}) + \cdots\,,
\end{split}
\end{equation}
where all the odd powers in $1/J_{6,\ell}$ cancel out upon using the large spin expansion of CFT data (\ref{eq: large_J_1}-\ref{eq: large_J_3}). We can therefore rewrite it in terms of TCBs as
\begin{eqnarray}
    &&\mathcal{F}^{(n)}= \label{eq: F_TCB_decomposition}\\
    &&\frac{L_z^n}{n!}\sum_{i}
    \begin{pmatrix}
    n\\ i
    \end{pmatrix}
    \left[
     \frac{\gamma_E^{n-i}}{2^i} H_{2}^{(0,i)} +\frac{n-i}{3}
     \frac{\gamma_E^{n-1-i}}{2^{i+1}} H_{2}^{(1,i)}
    \right]+\cdots\,.\nonumber
\end{eqnarray}
We have only showed the LL part for brevity and left the NLL part to Supplemental Material. Substituting in the TCBs using (\ref{eq: 4d_TCB}) and (\ref{eq: Hb_eg}) we obtain $\mathcal{F}^{(n)}$ at LP in $z$ and NLP in $1-\bar{z}$ 
\begin{equation}
\mathcal{F}^{(n)}=\frac{(-1)^n z^3}{2^{n+1} n!}L_z^n L_\epsilon^n \! \left[
\frac{1-\epsilon}{\epsilon} +\frac{2n}{3 L_\epsilon} + \frac{1}{L_z}+ \! \cdots \!
\right]+\mathcal{O}(z^4)\,,
\end{equation}
with NLL accuracy. Using $\mathcal{F}^{(n)}(z,\bar{z})=\frac{v}{u^3}\Phi^{(n)}(u,v)$ and crossing symmetry $\Phi(u,v)=\Phi(v,u)$, we get the NLL prediction for $\Phi^{(n)}$ at NLP in the double lightcone limit
\begin{eqnarray}\label{eq: Phi_NLL}
    && \Phi^{(n)} = \frac{(-1)^n}{2^n n!} L_u^n L_v^n\left\{
    \frac{1}{2}+(u+v)\right.\\
    &&\left.+\left[\left(\frac{n+1}{2}u+\frac{n}{3}v\right)\frac{1}{L_v}+(u\leftrightarrow v)\right]+\cdots
    \right\}  \,, \quad n>1 \nonumber\,.
\end{eqnarray}
As was promised in the last section, only twist-2 CFT data was used in the whole process. This prediction is checked against the available two- and three-loop results in the Supplemental Material.

\section{Power corrections to EEC in \texorpdfstring{$\mathcal{N} = 4$}{Lg} SYM}

The final task is to find the explicit relation between the double lightcone limit series $u^{j_1} v^{j_2} L_u^m L_v^n$ and the back-to-back limit series $y^j L_y^k$. The answer can be found using the Mellin representation of EEC~\cite{Belitsky:2013bja,Belitsky:2013xxa,Belitsky:2013ofa}
\begin{equation}
    \mathrm{EEC}(y)=\frac{1}{4y(1-y)^2}\int \frac{dj_1 dj_2}{(2\pi i)^2} M(j_1,j_2) K(j_1,j_2;y)\,,
\end{equation}
where $M(j_1,j_2)$ is the Mellin amplitude for $\Phi(u,v)$: $\Phi(u,v)=\int\frac{dj_1 dj_2}{(2\pi i)^2} M(j_1,j_2) u^{j_1}v^{j_2}$ and the kernel $K(j_1, j_2;y)$ is defined as
\begin{equation}
K(j_1,j_2;y)=\frac{2\Gamma(1-j_1-j_2) \left(\frac{y}{1-y}\right)^{j_1+j_2}}{\Gamma(j_1+j_2)\left[\Gamma(1-j_1)\Gamma(1-j_2)\right]^2}\,.
\end{equation}
This gives the map $u^{j_1}v^{j_2}\to  \frac{K(j_1,j_2;y)}{4y(1-y)^2}$. Taking derivatives w.r.t. $j_1$, $j_2$  generates the rules containing logarithms in $u$, $v$. One subtlety is that, due to the presence of the pole at $j_1+j_2=1$ in $K(j_1,j_2;y)$, the maps at NLP cannot predict the $y^0$ term without any $\log y$ enhancement. But it can be shown that all the logarithmic contributions at NLP are preserved~(see Supplemental Material). 
The rules at LP and NLP up to NLL are summarized in Table \ref{tab: log_map}, and lead to
\begin{align} 
\nonumber
 &\mathrm{EEC}^{(n>1)}(y) = \frac{ (-1)^{n} }{2^n (n-1)!} \left[ \frac{1}{2y} \left(
L_y^{2n-1} +  {\cal O}(L_y^{2n-3}) \right)
\right.
\\
 & \left. + \left( \frac{n}{2n-1} L_y^{2n-1} +\frac{7n-5}{12 } L_y^{2n-2} +{\cal O}(L_y^{2n-3}) \right)
  + \cdots
\right] \,, \label{eq:all-n-result}
\end{align}
which is in full agreement with the full theory calculation up to $n=3$ in \cite{Henn:2019gkr}. The $n>3$ terms are new and are one of the main result of this work. 
The analytic series in $n$ can be resummed explicitly to all orders, leading to the following NLL formula at LP and NLP\footnote{The one-loop NLL contribution at NLP has no $L_y$ enhancement. Therefore, we need to input the one-loop EEC to fix the constant $a/12$.}
\begin{eqnarray}
    \mathrm{EEC}(y)= -\frac{a L_y e^{-\frac{a L_y^2}{2}}}{4y} -\frac{1}{4}\left[
    \sqrt{\frac{\pi}{2}}\sqrt{a}\, \mathrm{erf}\left(\sqrt{\frac{a}{2}}L_y\right)\right.\nonumber\\
    \left. + a L_y e^{-\frac{a L_y^2}{2}}
    \right] + \frac{a}{48}(7 a L_y^2-4)e^{-\frac{a L_y^2}{2}} + \frac{a}{12} +\cdots\,,\quad \label{eq:resummed-res}
\end{eqnarray}
where $\mathrm{erf}$ is the error function $\mathrm{erf}(x)=\frac{2}{\sqrt{\pi}}\int_0^x e^{-t^2} dt$~\footnote{We note that the LL-NLP series has been studied in \cite{Moult:2019vou}. Their results disagree with ours starting from ${\cal O}(a^3)$, and seems to be in conflict with the fixed-order analytic result in \cite{Henn:2019gkr}. Further comparison is provided in the Supplemental Material.}. 

In Fig.~\ref{fig:my_label} we plot the ${\cal N} = 4$ EEC in the back-to-back limit to illustrate the importance of NLP resummation. It can be seen that the LL and NLL series at NLP leads to substantial corrections for not too large $\theta$. For $\theta > 175^{\circ}$ the Sudakov double logs suppressed the NLP contributions.  For comparison we also plot in dashed line the fixed-order NLP results truncated to NNLO~\cite{Henn:2019gkr}, along with the LP NLL series. In this case sizable NLP corrections can be found for $\theta > 175^{\circ}$, which however is misleading as they disappear after resumming to all orders in coupling. 

\begin{figure}
    \centering
    \includegraphics[width=\linewidth]{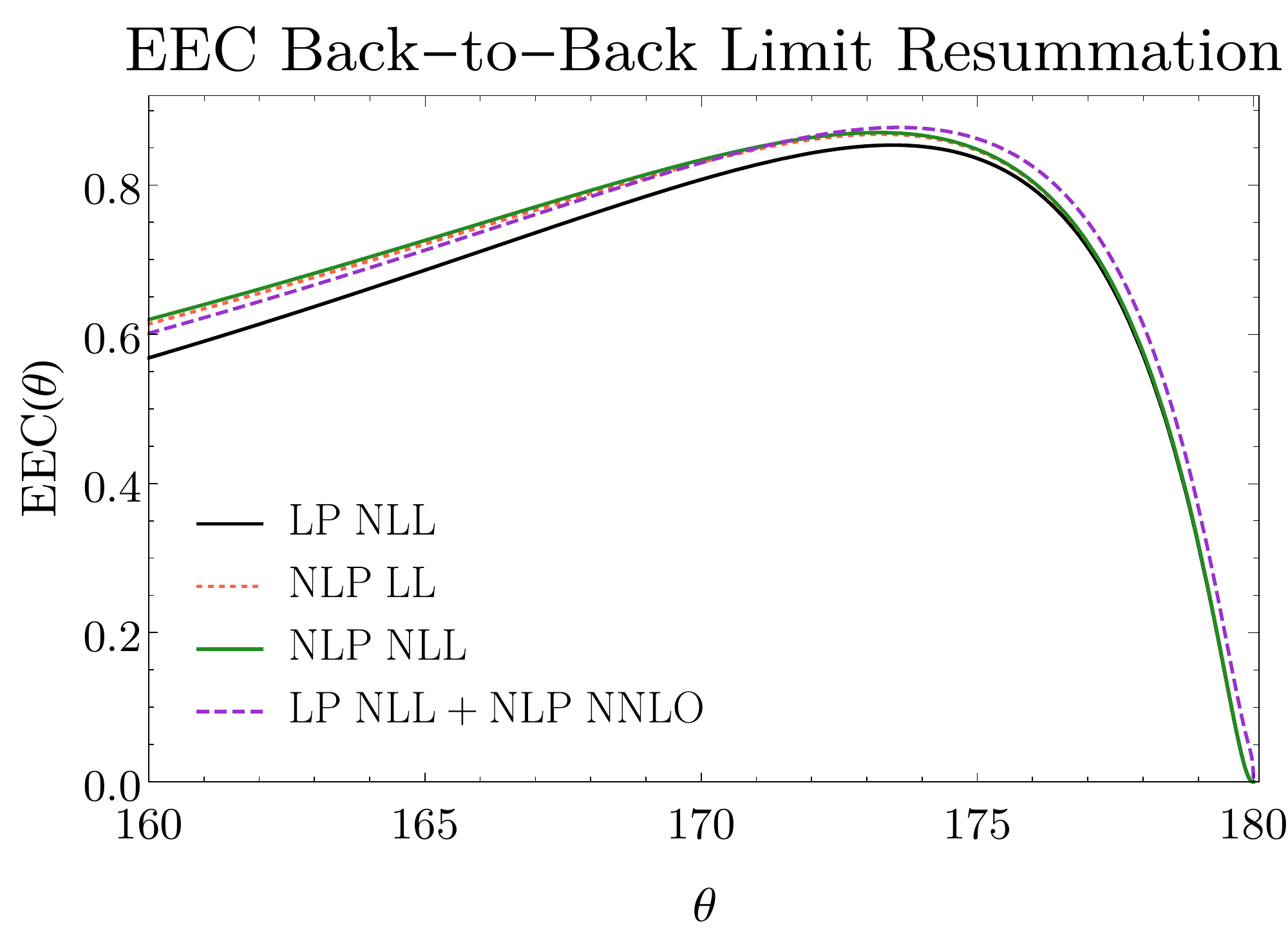}
    \caption{EEC as a function of $\theta$ in the back-to-back limit. We use $g^2/(4 \pi) = 0.118$ to mimic the QCD strong coupling at Z pole. The dashed line refers to LP resummed to NLL, with the inclusion of NLP terms up to NNLO~($n \leq 3$).}
    \label{fig:my_label}
\end{figure}

\section{Discussions}

 Our results lead to several exciting research avenues. First of all, it is interesting to apply the results to EEC in QCD, where fixed-order data up to NLO has become available  recently~\cite{Dixon:2018qgp,Luo:2019nig,Gao:2020vyx}. The local correlator of four electromagnetic currents in QCD has also been computed at one loop~\cite{Chicherin:2020azt}.  Secondly, in QCD running coupling corrections will modify NLL series. It would be important to understand how to incorporate these effects while retaining the power of conformal symmetry. Thirdly, our results provide concrete data for quantitative comparison between additive and multiplicative scheme in resummation matching, see {\it e.g.} \cite{Bizon:2018foh}. Fourthly, local correlators exhibit other interesting limits, such as the Regge limit~\cite{Costa:2012cb}. It would be interesting to understand what constraints are imposed on EEC by such limits. Last but not least, it would be worthwhile to understand the relation between our approach and the conventional approach based on momentum space renormalization group, in particular the relation between crossing symmetry for local correlator and the consistency relations from infrared poles cancellations~\cite{Moult:2016fqy}.

\begin{acknowledgments}
We thank Zhongjie Huang, Kai Yan, and Xiaoyuan Zhang for useful discussions. H.C. and H.X.Z. are supported by the Natural Science Foundation of China under contract No.~11975200 and No.~12147103. X.N.Z. is supported by funds from UCAS and KITS, and by the Fundamental Research Funds for the Central Universities.
\end{acknowledgments}

\bibliography{EEC}

\newpage

\onecolumngrid
\newpage
\appendix

\renewcommand{\theequation}{S.\arabic{equation}}
\setcounter{equation}{0}

\section*{Supplemental material}

\section{Scalar Local Correlator}
To make precise the local correlator in (\ref{4ptfun}), we consider the following  operator made out of the six scalars $\phi^{I=1,\ldots,6}$ of $\mathcal{N}=4$ SYM
\begin{equation}
O^{IJ}=\mathrm{tr} \left(\phi^I \phi^J\right)-\frac{1}{6}\delta^{IJ} \mathrm{tr}\left(\phi^K \phi^K \right)\,,
\end{equation}
The operator is the superprimary of the stress tensor multiplet and transforms in the symmetric traceless representation of the $SO(6)$ R-symmetry group. It is convenient to keep track of the R-symmetry information by contracting the indices with null polarization vectors $Y_I$ or with a traceless symmetric tensor $S_{IJ}$\footnote{The tensor can be built from the vectors, {\it e.g.}, $S_{IJ}=Y'_IY'_J+Y''_IY''_J$. Therefore, it is sufficient to focus on the former case.}
\begin{equation}
O(x,Y)\equiv O^{IJ}(x) Y_I Y_J\,, \quad \text{or} \quad O(x,S)=O^{IJ}(x) S_{IJ}\,.
\end{equation}
Due to superconformal symmetry, the four-point function has following ``partially non-renormalized'' form \cite{Eden:2000bk}
\begin{equation}\label{scalarcorrelator}
\langle O(x_1,Y_1) O(x_2,Y_2) O(x_3,Y_3) O(x_4,Y_4) \rangle = G^{(0)}(1,2,3,4)+\frac{2(N_c^2-1)}{(4\pi^2)^4} \frac{y_{12}^4y_{34}^4}{x_{12}^2x_{34}^2x_{14}^2x_{23}^2} R\; \Phi(u, v)
\end{equation}
where $G^{(0)}(1,2,3,4)$ is the tree-level correlator 
\begin{eqnarray}
G^{(0)}(1,2,3,4)=&&
\frac{(N_c^2-1)^2}{4(4\pi^2)^4}\left(
\left(\frac{y_{12}^2 y_{34}^2}{x_{12}^2 x_{34}^2}\right)^2
+\left(\frac{y_{13}^2 y_{24}^2}{x_{13}^2 x_{24}^2}\right)^2
+\left(\frac{y_{41}^2 y_{23}^2}{x_{41}^2 x_{23}^2}\right)^2
\right)\nonumber\\
&&+\frac{N_c^2-1}{(4\pi^2)^4}\left(
\frac{y_{12}^2 y_{23}^2 y_{34}^2 y_{41}^2}{x_{12}^2 x_{34}^2 x_{23}^2 x_{41}^2}
+\frac{y_{12}^2 y_{24}^2 y_{34}^2 y_{13}^2}{x_{12}^2 x_{24}^2 x_{34}^2 x_{31}^2}
+\frac{y_{13}^2 y_{23}^2 y_{24}^2 y_{41}^2}{x_{13}^2 x_{23}^2 x_{24}^2 x_{41}^2}
\right)\,,
\end{eqnarray}
with $y_{ij}^2\equiv Y_i\cdot Y_j$ and the function $\Phi(u,v)$ encodes all the dynamical information. The factor $R$ is determined by superconformal symmetry
\begin{equation}
    R=(1-z\alpha)(1-z\bar{\alpha})(1-\bar{z}\alpha)(1-\bar{z}\bar{\alpha})\;,
\end{equation}
where the conformal cross ratios $z$, $\bar{z}$ have already been introduced in the main text and $\alpha$, $\bar{\alpha}$ are similarly the R-symmetry cross ratios defined by
\begin{equation}
\frac{y_{13}^2y_{24}^2}{y_{12}^2y_{34}^2}=\alpha\bar{\alpha}\;,\quad \frac{y_{14}^2y_{23}^2}{y_{12}^2y_{34}^2}=(1-\alpha)(1-\bar{\alpha})\;.
\end{equation}
The weak coupling $g\ll 1$ expansion of $\Phi(u,v)$ reads 
\begin{equation}
\Phi(u,v) = \sum_{n=1}^{\infty} a^n \Phi^{(n)}(u,v)\,,\quad\quad  a=\frac{g^2 N_c}{4\pi^2}\;,
\end{equation}
and is known up to three loops \cite{Drummond:2013nda}. 
Since the dynamic function $\Phi(u,v)$ is R-symmetry independent, we can choose special polarizations to simplify the correlator (\ref{scalarcorrelator}). Following  \cite{Belitsky:2013xxa}, let us take 
\begin{equation}\label{eq: R_projector}
Y_I=(1,0,1,0,i,i)\,,\quad S_{IJ}=\mathrm{diag}(1,-1,0,0,0,0)\,, \quad S^\prime_{IJ}=\mathrm{diag}(0,0,1,-1,0,0)\,,
\end{equation}
and define 
\begin{equation}
\begin{split}
&\widetilde{\mathcal O}(x_2) = 2 O(x_2, S)\,,\quad
\widetilde{\mathcal O}^\prime (x_4) = 2 O(x_4, S^\prime)\,,\\
&  \mathcal{O}(x_3)=\left(\frac{N_c^2-1}{2\pi^4}\right)^{-\frac{1}{2}}O(x_3, Y)\,,\quad \mathcal{O}^\dagger(x_1) = \left(\frac{N_c^2-1}{2\pi^4}\right)^{-\frac{1}{2}}O(x_1, Y^*)\,.
\end{split}
\end{equation}
Then the four-point function becomes 
\begin{equation}
\begin{split}
\langle \mathcal{O}^\dagger (x_1) \widetilde{\mathcal O}(x_2) \widetilde{\mathcal O}^\prime (x_4) \mathcal{O}(x_3)\rangle
={}& \frac{1}{(2\pi)^4}\frac{1}{(x_{12}^2 x_{34}^2)^2}\left[\frac{N_c^2-1}{8}\left(1+\frac{u^2}{v^2}\right)
+\frac{u}{v}\left(\frac{1}{2}+\Phi(u,v)\right)
\right]\\
={}&\frac{1}{(2\pi)^4}\frac{1}{(x_{12}^2 x_{34}^2)^2}\left(\frac{N_c^2-1}{8} G^{\text{short}}(u,v)+\frac{1}{u^2}\mathcal{F}(u,v)\right)
\;,\end{split}
\end{equation}
where we have further split it into short multiplet contribution $G^{\text{short}}(u,v)$ and the long multiplet contribution $\mathcal{F}(u,v)$. The explicit form of $G^{\text{short}}(u,v)$ can be found in \cite{Beem:2016wfs} and is protected from perturbative corrections. As a result, $\Phi(u,v)$ is essentially the same as all the loop corrections ($n\geq 1$) of $\mathcal{F}(u,v)=\sum_{n\geq 0} a^n \mathcal{F}^{(n)}(u,v)$, {\it i.e.},
 \begin{equation}
 \mathcal{F}(u,v)-\mathcal{F}^{(0)}(u,v)=\frac{u^3}{v}\Phi(u,v)\,.
 \end{equation}
 Another reason for choosing the polarizations (\ref{eq: R_projector}) is the relation to the scalar detectors 
 \begin{equation}
\mathcal{S}(n_i)=\frac{1}{4}\int \displaylimits_{-\infty}^\infty dn_i\cdot x_i \lim_{\bar{n}_i\cdot x_i \to \infty} (\bar{n}_i\cdot x_i)^2 \mathcal{O}(x_i)\,.
 \end{equation}
Thanks to superconformal symmetry, the spinning correlator $\langle JTTJ\rangle$ is related to the scalar correlator (\ref{scalarcorrelator}) by Ward identities. Moreover, the EEC and the scalar-scalar correlation (SSC) are also proportional \cite{Belitsky:2013xxa}
\begin{equation}\label{eq: SSC_to_EEC}
\langle \mathcal{E}(n_2)\mathcal{E}(n_4) \rangle =\frac{4 (q^2)^2}{(n_2\cdot n_4)^2} \langle \mathcal{S}(n_2) \mathcal{S}(n_4)\rangle\,.
\end{equation}

\section{Twist Conformal Blocks}

The TCBs with logarithms can be computed using the method of \cite{Henriksson:2017eej}. The idea is to apply  the recursion relations on $\bar{H}_{\tau_0}^{(0,0)}(\bar{z})$, which is determined by the tree-level correlator, and compute $\bar{H}_{\tau_0}^{(m,0)}(\bar{z})$ at negative integer $m$. We then analytic continue in $m$ and take derivatives to obtain the logarithms. For our case, $\tau_0=2$ and we have
\begin{equation}
\bar{H}_{2}^{(0,0)}(\bar{z})= \frac{\bar{z}^2 (2-\bar{z})}{2(1-\bar{z})} + \bar{z} \log(1-\bar{z}) = \frac{1}{2\epsilon} + \text{regular terms }\,.
\end{equation}
Repeated $\bar{D}$ action gives $\bar{H}_2^{(m,0)}(\bar{z})$ and the first few terms in small $\epsilon$ for negative integer $m$ are given by (3.37) of ~\cite{Henriksson:2017eej} 
\begin{equation}
\begin{split}
\bar{H}^{(m,0)}_{2}(\bar{z}) = & 
\frac{1}{2} \epsilon^{m-1} \Gamma (1-m)^2+\frac{1}{6} m \left(2 m^2-6 m+1\right) \epsilon^m \Gamma (-m)^2\\
&+\frac{1}{180} (m-1) m (m+1) \left(20 m^3-54 m^2-35 m+36\right) \epsilon^{m+1} \Gamma (-m-1)^2
+\cdots \,.
\end{split}
\end{equation}
Obtaining the full analytic expression for $\bar{H}^{(m,i)}_{2}(\epsilon) $ is difficult. But life is much easier if we content ourselves with getting first few orders in $\epsilon$ and logarithms. Truncated to order $\epsilon^0$, only $\bar{H}^{(0,i)}_{2}(\epsilon) $ and $\bar{H}^{(1,i)}_{2}(\epsilon) $ are relevant and we find
\begin{eqnarray} \label{eq: Hb_LP}
\bar{H}^{(0,i)}_{2}(\bar{z}) &=& \frac{(-1)^i}{\epsilon} \left[
\frac{1}{2}\log^i\epsilon + i\gamma_E L^{i-1}_\epsilon+\frac{i(i-1)}{12}(12\gamma_E^2+\pi^2)L^{i-2}_\epsilon+\cdots
\right]\\
&+&  (-1)^i \left[
\frac{1}{6(i+1)}L^{i+1}_\epsilon+\frac{\gamma_E-3}{3}L^i_\epsilon+
\frac{\pi^2+12\gamma_E^2-72\gamma_E+12}{36}i L^{i-1}_\epsilon+\dots
\right]+\cdots \,, \nonumber \\
\bar{H}^{(1,i)}_{2}(\bar{z}) &=& 
(-1)^i\left[
\frac{1}{2(i+1)(i+2)}L^{i+2}_\epsilon
+\frac{\gamma_E}{i+1}L^{i+1}_\epsilon
+\frac{12\gamma_E^2+\pi^2}{12}L^i_\epsilon+\cdots
\right] +\cdots\,. \label{eq: Hb_NLP}
\end{eqnarray}

\section{More Details of (\ref{eq: F_twist_NLL_expand}) and (\ref{eq: F_TCB_decomposition})}
In this section, we present the details of the large spin perturbation calculation needed for obtaining EEC in the back-to-back limit to the NLL and NLP order. As we explained in the main text, only twist-2 contributions are needed. The expansion of the twist-2 conformal block $\mathbf{G}_{2+\gamma_{2,\ell},\ell}(z,\bar{z})$ in the $z\to 0$ limit is 
\begin{equation}
\begin{split}
\mathbf{G}_{2+\gamma_{2,\ell},\ell}(z,\bar{z}) 
&= z^{3+\gamma_{2,\ell}/2}k_{6+2\ell+\gamma_{2,\ell}}(\bar{z}) + \mathcal{O}(z^4) \\
&=z^3 \sum_{n=0}^{\infty} a^n \Bigg\{ \left[ \frac{1}{2^n n!}\left(\gamma^{(1)}_{2,\ell}\right)^n\log^n z
+ \frac{1}{2^{n-1}(n-2)!}\gamma^{(2)}_{2,\ell}\left(\gamma^{(1)}_{2,\ell}\right)^{n-2}  \log^{n-1}z + \cdots
\right] k_{6+2\ell}(\bar{z})\\
&\qquad\qquad\qquad + \frac{1}{2^{n} (n-1)!} \left(\gamma^{(1)}_{2,\ell}\right)^{n} \log^{n-1} z \;\partial_{\ell} k_{6+2\ell}(\bar{z}) +\cdots \Bigg\} + \mathcal{O}(z^4)\,.
\end{split}
\end{equation}
Combined with the  expansion of the OPE coefficient $a_{2,\ell}$, we obtain the leading twist contribution to $\mathcal{F}(z,\bar{z})$ 
\begin{equation} \label{eq: long_F_NLL_1}
\begin{split}
\mathcal{F}^{(n)}(z,\bar{z})
&= z^3 \sum_{\text{even }\ell} \Bigg\{ \log^n z  \frac{1}{2^n n!} a^{(0)}_{2,\ell} \left(\gamma^{(1)}_{2,\ell}\right)^n k_{6+2\ell}(\bar{z}) + \log^{n-1} z 
\Bigg[ a^{(1)}_{2,\ell} \frac{\left(\gamma^{(1)}_{2,\ell}\right)^{n-1}}{2^{n-1}(n-1)!}  
+ a^{(0)}_{2,\ell}  \frac{\gamma^{(2)}_{2,\ell}\left(\gamma^{(1)}_{2,\ell}\right)^{n-2} }{2^{n-1}(n-2)!}
\Bigg] k_{6+2\ell}(\bar{z})\\
&\qquad + \log^{n-1} z \;
a^{(0)}_{2,\ell}\frac{1}{2^{n} (n-1)!} \left(\gamma^{(1)}_{2,\ell}\right)^{n} \partial_{\ell} k_{6+2\ell}(\bar{z}) +\cdots 
\Bigg\} + \mathcal{O}(z^4)\,,
\end{split}    
\end{equation}
which is NLL in $\log z$. Then using the IBP identity
\begin{equation} \label{eq: IBP_id}
a^{(0)}_{2,\ell} \partial_{\ell} k_{6+2\ell}(\bar{z})= \partial_{\ell} [a^{(0)}_{2,\ell} k_{6+2\ell}(\bar{z})] - k_{6+2\ell}(\bar{z}) \partial_{\ell} a^{(0)}_{2,\ell}\,,
\end{equation}
and $a^{(1)}_{2,\ell}=-\zeta_2 a^{(0)}_{2,\ell} + \frac{1}{2} \partial_{\ell} \left(a^{(0)}_{2,\ell} \gamma^{(1)}_{2,\ell}\right)$, we rewrite (\ref{eq: long_F_NLL_1}) as
\begin{equation}
\begin{split}
\mathcal{F}^{(n)}(z, \bar{z}) =
& z^3 \sum_{\text{even }\ell}\left\{ \frac{1}{2^n n!}   \log^n z\;  a^{(0)}_{2,\ell} \left(\gamma^{(1)}_{2,\ell}\right)^n k_{6+2\ell}(\bar{z}) \right.+  \frac{1}{2^{n} (n-1)!}   \log^{n-1} z \, \bigg[
\left(\gamma^{(1)}_{2,\ell}\right)^{n} \partial_{\ell}\left(a^{(0)}_{2,\ell} k_{6+2\ell}(\bar{z})\right)\\
&\qquad\qquad \left.
+ a^{(0)}_{2,\ell} k_{6+2\ell}(\bar{z}) \left(\gamma^{(1)}_{2,\ell}\right)^{n-2}
\left[ \left(-2\zeta_2+\partial_\ell \gamma^{(1)}_{2,\ell}\right) \gamma^{(1)}_{2,\ell}
+ 2(n-1) \gamma^{(2)}_{2,\ell} \right] \bigg] +\cdots \right\}+\mathcal{O}(z^4)\,.
\end{split}    
\end{equation}
The use of (\ref{eq: IBP_id}) becomes clear when we use the integer-step finite difference to approximate $\partial_{\ell}\left(a^{(0)}_{2,\ell} k_{6+2\ell}(\bar{z})\right)$ at large spin $\ell$. On a general function $f(\ell)$, we approximate $f^\prime(\ell) \approx \frac{f(\ell+2)-f(\ell-2)}{4}$, which is accurate up to $\mathcal{O}(\ell^{-2})$. Neglecting boundary terms which vanish at large spins, we can write   
\begin{equation}
\sum_{\text{even }\ell} 
\left(\gamma^{(1)}_{2,\ell}\right)^{n} \partial_{\ell}\left(a^{(0)}_{2,\ell} k_{6+2\ell}(\bar{z})\right)
\approx \frac{1}{4} \sum_{\text{even }\ell} a^{(0)}_{2,\ell} k_{6+2\ell}(\bar{z})
\left[\left(\gamma^{(1)}_{2,\ell-2}\right)^{n} - \left(\gamma^{(1)}_{2,\ell+2}\right)^{n} \right]\,,   \end{equation}
Expanding everything other than $a_{2,\ell}^{(0)}$ with respect to the large conformal spin $J_{6,\ell}^2$, we get 
\begin{equation}
\begin{split}
\mathcal{F}^{(n)}(z, \bar{z}) =
 z^3  \sum_{\text{even }\ell}& \left\{ \frac{1}{2^n n!}   \log^n z \; a^{(0)}_{2,\ell} k_{6+2\ell}(\bar{z}) \left[ \left(\log(J_{6,\ell}^2)+2\gamma_E\right)^n
 +\frac{n}{3} \left(\log(J_{6,\ell}^2)+2\gamma_E\right)^{n-1}\frac{1}{J_{6,\ell}^2}+\cdots\right] \right.\\
& + \frac{1}{2^{n} (n-1)!}  \log^{n-1} z \; a^{(0)}_{2,\ell} k_{6+2\ell}(\bar{z})  \bigg[ -(n+1)\zeta_2 \left(\log(J_{6,\ell}^2)+2\gamma_E\right)^{n-1}\\
& -3(n-1)\zeta_3 \left(\log(J_{6,\ell}^2)+2\gamma_E\right)^{n-2}
+\frac{1}{J^2_{6,\ell}}\bigg( (n-1)\left(1-\frac{\zeta_2}{3}(n+1)\right) \left(\log(J_{6,\ell}^2)+2\gamma_E\right)^{n-2}\\
&\left.
\hspace{4cm}-(n-1)(n-2)\zeta_3 \left(\log(J_{6,\ell}^2)+2\gamma_E\right)^{n-3}\bigg) \bigg] +\cdots
\right\}+\mathcal{O}(z^4)\,,
\end{split}    
\end{equation}
which can be organized into TCBs as
\begin{equation} \label{eq: app_TCB_decomp}
\begin{split} 
\mathcal{F}^{(n)}(z, \bar{z}) =&  
\frac{\log^n z}{2^n n!}
\left(\sum_{i=0}^{n}\frac{n!}{i ! (n-i)!} (2\gamma_E)^i H_{2}^{(0,n-i)}
+\frac{n}{3} \sum_{i=0}^{n-1-i} \frac{(n-1)!}{i! (n-1-i)!} (2\gamma_E)^i H_{2}^{(1,n-1-i)} +\cdots\right) \\
& +  \frac{\log^{n-1} z}{2^n (n-1)!}\bigg\{
-(n+1)\zeta_2 \sum_{i=0}^{n-1} \frac{(n-1)!}{i!(n-1-i)!} (2\gamma_E)^i H_{2}^{(0,n-1-i)} \\
&\qquad\qquad -3(n-1)\zeta_3 \sum_{i=0}^{n-2} \frac{(n-2)!}{i!(n-2-i)!} (2\gamma_E)^i H_{2}^{(0,n-2-i)}\\
&\qquad\qquad +(n-1)\left(1-\frac{\zeta_2}{3}(n+1)\right) \sum_{i=0}^{n-2} \frac{(n-2)!}{i! (n-2-i)!} (2\gamma_E)^i H_{2}^{(1,n-2-i)}\\
&\qquad\qquad -(n-1)(n-2)\zeta_3 \sum_{i=0}^{n-3}\frac{(n-3)!}{i!(n-3-i)!} (2\gamma_E)^i H_{2}^{(1,n-3-i)} +\cdots \bigg\} + \cdots\,.
\end{split}
\end{equation}
Substituting the explicit TCBs (\ref{eq: Hb_LP}, \ref{eq: Hb_NLP}), we get 
\begin{eqnarray}
\mathcal{F}^{(n)}(z,\bar{z})
&&=  z^3 \left\{ \frac{1}{n!} \log^n z \left[
\frac{1}{\epsilon}\left(\frac{(-1)^n}{2^{n+1}}\log^n \epsilon+\cdots\right)
+\left(\frac{(-1)^{n+1}}{2^{n+1}}\log^n\epsilon + \frac{(-1)^n n}{3\times 2^n}\log^{n-1}\epsilon+\cdots\right)
+\cdots\right]
\right. \nonumber\\
&&\qquad\quad + \frac{ \log^{n-1}z}{(n-1)!}\bigg[
\frac{1}{\epsilon}\left( \frac{(-1)^n }{2^{n+1}}(n+1)\zeta_2\log^{n-1}\epsilon
-\frac{(-1)^n}{2^{n+1}} 3(n-1)\zeta_3 \log^{n-2}\epsilon+\cdots \right) \nonumber\\
&&\left.
\hspace{3.5 cm} + \left(\frac{(-1)^n}{2^{n+1} n} \log^n\epsilon + \frac{(-1)^{n+1}}{2^{n+1}}(n+1)\zeta_2 \log^{n-1}\epsilon+\cdots\right) \bigg] +\cdots
\right\} +\mathcal{O}(z^4)
\,.
\end{eqnarray}
Via $\mathcal{F}^{(n)}(z,\bar{z})=\frac{v}{u^3}\Phi^{(n)}(z,\bar{z})$, this gives the expansion of $\Phi^{(n)}(z,\bar{z})$. Crossing symmetry allows us to further reconstruct the $\mathcal{O}(u^1 v^0)$ contributions, which gives the results in \eqref{eq: Phi_NLL}.

\section{Details of the Map from \texorpdfstring{$u^{j_1} v^{j_2} L_u^m L_v^n$}{Lg} to \texorpdfstring{$y^j L_y^k$}{Lg}}

In this section, we provide more details for establishing the map from the small $u,v$ expansion of $\Phi(u,v)$ to the small $y$ expansion of $\mathrm{EEC}(y)$. Instead of electromagnetic current sources $J_\mu$, we consider two scalar operator sources, belonging to the stress tensor multiplet, in the center of mass frame $q^\mu=(Q,0,0,0)$. $\mathrm{EEC}(y)$ relates to $\langle \mathcal{E}(n_2)\mathcal{E}(n_4)\rangle$ by an overall factor:
\begin{equation}
\mathrm{EEC}(y)=\int d\Omega_2 d\Omega_4 \delta(\vec{n}_2\cdot\vec{n}_4 -\cos \theta) \frac{\langle \mathcal{E}(n_2)\mathcal{E}(n_4)\rangle}{Q^2} =\frac{8\pi^2}{ Q^2} \langle \mathcal{E}(n_2)\mathcal{E}(n_4) \rangle\,,
\end{equation}
where $\theta$ is the angle between $\vec{n}_2$ and $\vec{n}_4$ and we assume the convention that $\langle \mathcal{E}(n_2)\mathcal{E}(n_4)\rangle$ has already been normalized to the cross section. The superconformal Ward identities further reduce the EEC $\langle \mathcal{E}(n_2)\mathcal{E}(n_4)\rangle$ to scalar-scalar correlation (SSC) $\langle \mathcal{S}(n_2) \mathcal{S}(n_4)\rangle$ ~\cite{Belitsky:2013xxa, Belitsky:2013ofa}
\begin{equation} 
\langle \mathcal{E}(n_2)\mathcal{E}(n_4) \rangle =\frac{4 (q^2)^2}{(n_2\cdot n_4)^2} \langle \mathcal{S}(n_2) \mathcal{S}(n_4)\rangle\,.
\end{equation}
The SSC is related to the local correlator in a simple way in Mellin space \cite{Belitsky:2013xxa}
\begin{equation}
\Phi(u,v)=\int \frac{dj_1 dj_2}{(2\pi i)^2} M(j_1,j_2) u^{j_1} v^{j_2}\;.
\end{equation}
Here $M(j_1,j_2)$ is the Mellin amplitude and encodes all the dynamical information. To compute the SSC, the first step 
is to obtain the Lorentzian correlator. This is achieved by using the Wightman prescription $x_{ij}^2\to -x_{ij}^2 + i\epsilon t_{ij}$, if operator $i$ sits before $j$. In our case, the operator ordering is $1<2<4<3$. The second step is to perform the light transform on the Lorentzian correlator
\begin{equation}
\prod_{i=2,4} \int_{-\infty}^{\infty}d(n_i\cdot x_i) \lim_{\bar{n}_i\cdot x_i \to \infty} \left(\frac{\bar{n}_i\cdot x_i}{2}\right)^2\,,
\end{equation}
which turns local operators into detectors.
To obtain SSC in the momentum space, the last step is the Fourier transformation $\int d^4 x_{13} \exp(i q\cdot x_{13})$. After normalized to the total cross section, the Mellin representation for SSC is
\begin{equation}
\langle \mathcal{S}(n_2)\mathcal{S}(n_4) \rangle
=\frac{1}{(2\pi)^4}\frac{\pi^2}{(n_2\cdot n_4) q^2} \frac{1-y}{y}\int \frac{dj_1 dj_2}{(2\pi i)^2} M(j_1, j_2) K(j_1,j_2;y)\,,
\end{equation}
with
\begin{equation}
K(j_1,j_2;y)=\frac{2\Gamma(1-j_1-j_2)}{\Gamma(j_1+j_2)\left[\Gamma(1-j_1)\Gamma(1-j_2)\right]^2} \left(\frac{y}{1-y}\right)^{j_1+j_2}\,.
\end{equation}
Then using $n_2\cdot n_4=2(1-y)$ in the center of mass frame, we obtain the Mellin representation for $\mathrm{EEC}(y)$
\begin{equation}
\mathrm{EEC}(y)=\frac{8\pi^2 Q^2}{(1-y)^2}\langle \mathcal{S}(n_2)\mathcal{S}(n_4)\rangle = \frac{1}{4 y(1-y)^2} \int \frac{dj_1 dj_2}{(2\pi i)^2} M(j_1, j_2) K(j_1, j_2;y)\,.
\end{equation}
Therefore, we expect the following mapping from the double lightcone limit to back-to-back limit
\begin{equation}
    u^{j_1} v^{j_2} \to \frac{K(j_1,j_2;y)}{4 y(1-y)^2 }\,.
\end{equation}
At LP and NLP, we need the following rules containing logarithms in $u,v$
\begin{eqnarray}
\log^m u \log^n v &\;\to\;& m!\, n!\oint_{|j_1|=\epsilon} \frac{dj_1}{2\pi i}\frac{1}{j_1^{m+1}} \oint_{|j_2|=\epsilon} \frac{dj_2}{2\pi i}\frac{1}{j_2^{n+1}} \frac{K(j_1,j_2;y)}{4y(1-y)^2}\,,  \label{eq: abstract_map_LP}\\
u \log^m u \log^n v &\;\to\;& {m!\, n!}\oint_{|j_1-1|=\epsilon} \frac{dj_1}{2\pi i}\frac{1}{(j_1-1)^{m+1}} \oint_{|j_2|=\epsilon} \frac{dj_2}{2\pi i}\frac{1}{j_2^{n+1}}  \frac{K(j_1,j_2;y)}{4y(1-y)^2}\,,\label{eq: abstract_map_NLP_1}\\
v \log^m u \log^n v &\;\to\;& {m!\, n!}\oint_{|j_1|=\epsilon} \frac{dj_1}{2\pi i}\frac{1}{j_1^{m+1}} \oint_{|j_2-1|=\epsilon} \frac{dj_2}{2\pi i}\frac{1}{(j_2-1)^{n+1}} \frac{K(j_1,j_2;y)}{4y(1-y)^2} \,. \label{eq: abstract_map_NLP_2}
\end{eqnarray}
For the cases we will inspect, the only exception is the constant term at NLP which is caused by the $j_1+j_2=1$ pole in $K(j_1,j_2;y)$. To see this, we consider the first derivative of $\frac{1-y}{y}K(j_1,j_2;y)$ w.r.t. $\frac{y}{1-y}$. Such an action shifts $\Gamma(1-j_1-j_2)$ to $\Gamma(2-j_1-j_2)$, which is analytic at $j_1+j_2=1$. The constant term is killed after taking the derivative, while the logarithmic information remains. The explicit expressions for general $m,n$ up to NLL accuracy are shown in Table \ref{tab: log_map}.

\section{Comparison with Existing Results}

In this section we provide a comparison of our predictions with the existing results in the literature. We begin with the local correlator in the double lightcone limit to NLL accuracy. The full three-loop correlator can be found in \cite{Drummond:2013nda}\footnote{There is an overall normalization difference $\left(-\frac{1}{4\pi}\right)^n$ at the $n$-th loop.}. Upon expanding their results in the double lightcone limit we find 
\begin{eqnarray}
    \Phi^{(1)}(u,v) &=& \left[-\frac{1}{4}\log u\log v 
 + 0 \cdot \log (u v) +\cdots \right]
    - {\color{red} \left[\frac{1}{4}(u+v)\log u \log v+\frac{1}{2}(u\log u +v\log v )+\cdots\right]}+\cdots\,,\nonumber\\
    \Phi^{(2)}(u,v) &=& \left[\frac{1}{16}\log^2 u \log^2 v + 0 \cdot \log u \log v \log (uv)  +  \cdots \right]+\left[\frac{1}{8}(u+v)\log^2 u \log^2 v\right. \nonumber\\
    &&\qquad \left. +\frac{3}{16}\log u \log v (u \log u+v\log v ) +{\color{red}\frac{1}{8}\log u \log v (v\log u +u\log v )}+\cdots \right]+\cdots\,, \nonumber\\
    \Phi^{(3)}(u,v) &=&\left[-\frac{1}{96}\log^3 u\log^3 v+ 0 \cdot \log^2 u \log^2 v \log(uv)  + \cdots \right]-\left[\frac{1}{48}(u+v)\log^3 u \log^3 v \right.\nonumber\\
    &&\left.\qquad 
    + \frac{1}{24} \log^2 u \log^2 v (u \log u+v\log v )
    +\frac{1}{48} \log^2 u \log^2 v (v\log u +u\log v )+\cdots
    \right]+\cdots\,. \label{eq:local_three_loop}
\end{eqnarray}
Compared with our NLL prediction (\ref{eq: Phi_NLL}), we find perfect agreement except for the NLP terms at one loop and a single NLL-NLP term at two loops~(shown in red explicitly). Such terms do not correspond to enhanced divergence in $v$ and hence cannot be captured by the large spin analysis. Moreover, the two-loop NLL-NLP mismatched term ${\color{red}\frac{1}{8}\log u \log v (v\log u +u\log v )}$ does not map to NLL-NLP term in EEC, as can be checked from Table \ref{tab: log_map}.

For EEC in ${\cal N}=4$ SYM results up to three loops with full $y$ dependence have been computed in \cite{Henn:2019gkr}, using the local correlator from \cite{Drummond:2013nda} as input. 
The back-to-back expansion of \cite{Henn:2019gkr} up to NLL at NLP is
\begin{eqnarray}
\mathrm{EEC}^{(1)} &=& -\frac{1}{4 y}\log y {\color{red} -\frac{1}{2}\log y } {\color{red}+ 0\cdot y^0 } + \mathcal{O}(y)\,,\nonumber \\
\mathrm{EEC}^{(2)} &=& \frac{1}{y}\left(\frac{\log^3 y}{8}+0\cdot \log^2 y+\cdots\right) +\left(\frac{\log^3 y}{6}+\frac{3}{16}\log^2 y+\cdots\right) + \mathcal{O}(y) \nonumber \\
\mathrm{EEC}^{(3)} &=& \frac{1}{y}\left(-\frac{\log^5 y}{32}+0\cdot\log^4 y+\cdots\right)+\left(-\frac{3\log^5 y}{80}-\frac{\log^4 y}{12}+\cdots\right) + \mathcal{O}(y)\,. \label{eq:explicit_results}
\end{eqnarray}
To obtain (\ref{eq:explicit_results}),
it is necessary to expand the following integral to NLP,
\begin{equation}
\mathrm{E} = \int_0^1 d\bar z \int_0^{\bar z} dt \frac{-1}{t (1-y - \bar z) + y \bar z } \left[ 
\frac{z \bar z}{1 - z - \bar z} P_1 + \frac{z^2 \bar z}{(1-z)^2 (1-z \bar z)} P_2
\right] \,,
\end{equation}
where $z = (1-y) t (t- \bar z)/(t (1-y - \bar z) + y \bar z)$ and $P_1$ and $P_2$ are lengthy combinations of weight-3 harmonic polylogarithms and can be found in  \cite{Henn:2019gkr}. The expansion of this integral begins from NLP and reads
\begin{equation}
\mathrm{E} = \frac{\log
   ^5 y}{60}+ 0 \cdot \log^4 y + \frac{1}{12} \zeta_2 \log ^3 y + \cdots
   \label{eq:elliptic}
\end{equation}
where we have neglected terms of ${\cal O}(\log^2 y)$ and beyond. The expansion of this integral was also performed in \cite{Moult:2019vou}, but the result reported in Eq.~(5.17) of \cite{Moult:2019vou} is larger than \eqref{eq:elliptic} by a factor of two.
Using \eqref{eq:elliptic} and expanding the remaining results in \cite{Henn:2019gkr} with the package HPL, we obtain $\mathrm{EEC}^{(3)}$ presented in \eqref{eq:explicit_results}.

Our results in \eqref{eq:all-n-result} for $n>1$ are in full agreement with $\mathrm{EEC}^{(2)}$ and $\mathrm{EEC}^{(3)}$ in \eqref{eq:explicit_results}. This provides a strong check for our results. Our large spin analysis does not expect to capture the NLP terms at one loop~(shown in red in \eqref{eq:explicit_results}), for the same reason as explained for local correlator.

We note that a LL study at NLP has also been performed in \cite{Moult:2019vou}, where an RG equation at NLP has been derived. After changing to our normalization, their predicted LL coefficient at NLP and three loop, given in Eq.~(5.21) of \cite{Moult:2019vou}, is $-1/30 \log^5 y$, which disagrees with our results in \eqref{eq:resummed-res}, as well as with the full results from \cite{Henn:2019gkr}.

\end{document}